\documentclass[showpacs,showkeys,10pt,preprintnumbers,nofootinbib,groupedaddress,superscriptaddress]{revtex4-1}

\usepackage{amsmath}
\usepackage[dvipdfmx]{graphicx}
\usepackage{epstopdf}
\usepackage{dcolumn}
\usepackage{bm}
\usepackage[normalem]{ulem}
\allowdisplaybreaks

\newcommand{\nn}{\nonumber}

\newcommand{\be}{\begin{eqnarray}}
\newcommand{\ee}{\end{eqnarray}}
\catcode`\@=11
\def\lsim{\mathrel{\mathpalette\@versim<}}
\def\gsim{\mathrel{\mathpalette\@versim>}}
\def\@versim#1#2{\vcenter{\offinterlineskip
\ialign{$\m@th#1\hfil##\hfil$\crcr#2\crcr\sim\crcr } }}
\catcode`\@=12

\newcommand{\Slash}[1]{{\ooalign{\hfil#1\hfil\crcr\raise.167ex\hbo
x{/}}}}

\def\thefootnote{\fnsymbol{footnote}}

\begin{document}

\title{
Multicomponent Dark Matter in Radiative Seesaw Model \\
and Monochromatic Neutrino Flux
}

\author{Mayumi \surname{Aoki}}
\email{mayumi@hep.s.kanazawa-u.ac.jp}
\affiliation{Institute for Theoretical Physics, Kanazawa University, Kanazawa 920-1192, Japan}

\author{Jisuke \surname{Kubo}}
\email{jik@hep.s.kanazawa-u.ac.jp}
\affiliation{Institute for Theoretical Physics, Kanazawa University, Kanazawa 920-1192, Japan}

\author{Hiroshi \surname{Takano}}
\email{hiroshi.takano@ipmu.jp}
\affiliation{Institute for Theoretical Physics, Kanazawa University, Kanazawa 920-1192, Japan}
\affiliation{
Kavli IPMU (WPI), University of Tokyo, Kashiwa, Chiba 277-8583, Japan}

\preprint{KANAZAWA-14-07}
\preprint{IPMU14-0136}


\begin{abstract}
We consider a two loop radiative seesaw model with an
exact $Z_2 \times Z'_2$ symmetry, which 
 can stabilize two or three dark matter particles. 
 The model is a simple extension of the inert scalar model of Ma, where
the  lepton-number violating mass term of the 
 inert scalar,
 which is required to be small for small neutrino masses,
  is generated at the one-loop level.
 The semi-annihilation processes of different dark matter particles,
which are present when there exist more than three different
dark matter particles, not only  play an 
important role for their relic densities but also are responsible
for the monochromatic neutrino lines
resulting  from the dark matter annihilation processes.
The  monochromatic neutrinos do not suffer from a chiral suppression, and 
we investigate the observational possibility 
of the monochromatic neutrino flux from the Sun.

 \end{abstract}
\setcounter{footnote}{0}
\def\thefootnote{\arabic{footnote}}
\maketitle

\section{Introduction}

Tiny neutrino  masses and the absence of dark matter (DM) candidates are  problems
of the standard model (SM), which can be overcome only by its  extension.
The tiny neutrino  masses can be explained by
 the seesaw mechanism \cite{Minkowski:1977sc}, which usually  requires 
an introduction of right-handed neutrinos with lepton-number
violating Majorana masses. However, for the tree-level seesaw mechanism
to work, an undesirable  hierarchy in the mass scale
or in the size of the Yukawa couplings has to be introduced:
To obtain neutrino  masses of $\mathcal{O}(0.1)$ eV in the type I seesaw
for instance, we need either
 large Majorana masses of Grand Unified Theory scale or very small
Yukawa couplings of  $\mathcal{O}(10^{-6})$ of  the right-handed neutrinos
with the left-handed ones if  the Majorana masses are $\mathcal{O}(1)$ TeV.
This unwelcome feature can be avoided if the neutrino masses are generated 
radiatively 
(\cite{Krauss:2002px}-\cite{Ma:2006km} for instance). 
With  an increasing number of loops, the hierarchy 
between the SM scale and the Majorana masses becomes milder, and in fact
the Majorana masses  can become  $\mathcal{O}(1)$ TeV 
 without making the Yukawa couplings very small.
The common feature of the radiative seesaw models is 
the  existence of an unbroken discrete symmetry, which forbids 
the appearance of  Dirac neutrino masses.
An important consequence of this unbroken symmetry, usually $Z_2$, 
is that the lightest $Z_2$ odd particle is stable 
and hence  can be a DM candidate with 
a mass of $\mathcal{O}(1)$ TeV.

In the one-loop radiative seesaw model of Ma \cite{Ma:2006km},
the lepton-number violating mass term of the 
 inert doublet scalar $\eta$ is
required to be very small to obtain small neutrino masses.
The mass term originates from a lepton-number violating quartic 
scalar coupling,
the "$\lambda_5$ coupling", which is
$\mathcal{O}(10^{-5})$ to obtain small neutrino masses with the Yukawa couplings of  $\mathcal{O}(0.01)$.
In this paper, we consider an extension of  the model such that
this  lepton-number violating mass, too, is generated radiatively.
Consequently, the seesaw mechanism occurs at the two-loop level
in the extended model \cite{Aoki:2013gzs}
(A similar idea has been proposed 
in an $E_6$ inspired model \cite{Ma:2007yx}.)
.
For this mechanism to work, we have to introduce 
a larger unbroken discrete symmetry, $Z_2\times Z_2$,
which implies that the model yields a
multicomponent DM system
\cite{Ma:2007yx}-\cite{Aoki:2012ub}. 
We emphasize that the 
multicomponent DM system is a consequence of the
 unbroken $Z_2\times Z_2$, which forbids the Dirac neutrino masses
and also the one-loop neutrino mass diagram.

In  \cite{Aoki:2013gzs} we have investigated 
 in the extended model the two-component DM
system consisting of a neutral component of $\eta$ and 
another real scalar $\chi$. 
We have found that the  $\chi$ DM  can cover 
the shortage of the relic density of the $\eta$  DM 
  for  the mass range 100 GeV $\lesssim m_\eta \lesssim$ 600 GeV \cite{Barbieri:2006dq,LopezHonorez:2006gr}. 
We were motivated by the desire to explain at the same time
a slight excess of the Higgs decay into two
$\gamma$'s  \cite{:2012gk} and 
the $135$ GeV $\gamma$-ray line
possibly observed at the  Fermi LAT
\cite{Ackermann:2012qk}
by the annihilation of the $\chi$ DM. 
Though it is not impossible to explain both $\gamma$ excesses,
we have to use a corner of the parameter space, which faces the border
of perturbation theory.
Moreover, the subsequent experimental 
searches  could not confirm 
these interesting $\gamma$ excesses \cite{ATLAS:2013mma,Gustafsson:2013fca}.

In this paper we consider the same model in the  parameter space
leading to a three-component DM system:
Our DM candidates are the lightest right-handed neutrino $N$,
and two real scalars $\phi_R$ and $\chi$.
The semi-annihilation processes of these DM particles 
have a considerable influence on their relic densities
\cite{Aoki:2012ub,D'Eramo:2010ep}, and 
 the monochromatic neutrino lines
can be produced  from the 
semi-annihilation process such as $\phi_R \chi\to N\nu$.
These  monochromatic neutrinos are not  chirally suppressed, and 
we analyze the observational prospect
of the monochromatic neutrino flux from the Sun. 
Semi-annihilations of DM particles can produce 
line spectra of neutral SM
particles, e.g. neutrinos \cite{Aoki:2012ub} and photons 
\cite{D'Eramo:2012rr},
and observations of such line spectra are
indications of a multicomponent DM Universe \footnote{
Semi-annihilation processes exist also in one-component DM systems when DM is a $Z_3$ charged particle 
\cite{D'Eramo:2010ep} or a vector boson \cite{Hambye:2008bq}.
}.

\section{Model}
\begin{table}
\caption{\footnotesize{The matter contents of 
the model and the corresponding quantum numbers. $Z_{2}
\times  Z'_2$ is  the unbroken discrete symmetry,
while the lepton number $L$ is softly broken by the $\phi$ mass.}}
\begin{center}
\begin{tabular}{|c|c|c|c|c|c|c|} 
\hline
field & statistics& $SU(2)_L$ & $U(1)_Y$ & $Z_2$ &$Z'_2$ & $L$	\\ \hline
$(\nu_{L},l_L)$	& F& $2$ 	& $-1/2$ 	& $+$ & $+$ & $1$\\ \hline
$l^c_R$ 		& F	& $1$ 	& $1$		& $+$ & $+$ &$-1$	\\ \hline
$N^c_R$		& F		& $1$		& $0$ 	& $-$ & $+$ &$0$	\\ \hline
$H=(H^+,H^0)$ 	& B& $2$ 	& $1/2$ 	& $+$ & $+$	 &$0$\\ \hline
$\eta=(\eta^+,\eta^0) 	$ 	& B& $2$ 	& $1/2$ 	& $-$ & $+$ &$-1$	\\ \hline
$ \chi $     	& B  & $1$    & $0$        & $+$ & $-$	 &$0$\\ \hline
$ \phi $   	& B  & $1$    & $0$        & $-$ & $-$	 &$1$\\ \hline
\end{tabular}
\end{center}
\label{contents}
\end{table}
Here we will briefly outline the model \cite{Aoki:2013gzs}, where we show
the matter content of the model  in Table I.
In addition to the matter content of the SM model,
we introduce the right-handed neutrino $N_R^c$, 
an $SU(2)_L$ doublet scalar $\eta$, 
and two SM singlet scalars $\chi$ and $\phi$.
 Note that the lepton number $L$ of $N_R^c$ is zero.
The $Z_2 \times Z'_2 \times L$ -invariant 
Yukawa sector and Majorana mass term for $N_R^c$ can be described by
\begin{equation}
\mathcal{L}_Y 
= Y^e_{ij} H^\dag L_i  l_{Rj}^c
 + Y^{\nu}_{ik}L_i \epsilon \eta N_{Rk}^c -\frac{1}{2} 
 M_{k} N_{Rk}^c N_{Rk}^c 
 + h.c. ~,
 \label{LY}
\end{equation}
where $i,j,k~(=1,2,3)$ stand for  the flavor indices.
The scalar potential $V$ is written as $V=V_{\lambda}+V_m$,
where 
\begin{eqnarray}
V_{\lambda}&=&
\lambda_1 (H^\dag H)^2
 +\lambda_2 (\eta^\dag \eta)^2
 + \lambda_3 (H^\dag H)(\eta^\dag \eta)
+\lambda_4 (H^\dag \eta)(\eta^\dag H)
 \nonumber \\
 & &
 + \gamma_1 \chi^4
 +\gamma_2 (H^\dag H)\chi^2
 + \gamma_3 (\eta^\dag \eta)\chi^2
 + \gamma_4 |\phi|^4
+ \gamma_5 (H^\dag H)|\phi|^2\nn\\
  & &+ \gamma_6 (\eta^\dag \eta)|\phi|^2
+ \gamma_7 \chi^2|\phi|^2
+ \frac{\kappa}{2} [\,(H^\dag \eta)\chi \phi+h.c.\,]~,
 \label{potential}\\
V_{m}&=&m_1^2 H^\dag H + m_2^2 \eta^\dag \eta 
+ \frac{1}{2} m_3^2 \chi^2+ m_4^2 |\phi|^2
+ \frac{1}{2} m_5^2 [\, \phi^2+(\phi^*)^2\,]~.
\label{v2A}
\ee
The potential $V$, except the last term in $V_m$,
 is $Z_2 \times Z'_2 \times L$ -invariant.
 This last term breaks the lepton number softly.
In the absence of this term, there will be no neutrino mass.
Note that  the ``$ \lambda_5$ term'', 
$(1/2)\lambda_5 (H^\dag \eta)^2$,  is also forbidden by $L$.
A small $\lambda_5$ of the original model of Ma \cite{Ma:2006km}
is ``natural'' according to 't Hooft \cite{'tHooft:1979bh}, because the absence of 
$\lambda_5$ implies an enhancement of symmetry. 
In fact, if $\lambda_5$ is small at some scale, it remains
small for other scales as one can explicitly verify \cite{Bouchand:2012dx}.
Here we attempt to derive the smallness of $\lambda_5$
dynamically, such that the $\lambda_5$ term becomes calculable.

The charged, CP even and odd scalars are defined as
\be
H &=& \left(\begin{array}{c}H^+ \\ (v_h+h+i G)/\sqrt{2}\\
\end{array}\right)~,~
\eta = \left(\begin{array}{c}\eta^+ 
\\ (\eta_{R}^0+i\eta_{I}^0)/\sqrt{2}\\
\end{array}\right)~,~\phi=(\phi_R+i \phi_I)/\sqrt{2}~,
\ee
where $v_h$ is the vacuum expectation value.
The tree-level masses of the scalars are given by
\begin{align}
m_h^2 &=2 \lambda_1 v_h^2~,\\
 m^2_{\eta^\pm}&=m_2^2+\frac{1}{2}\lambda_3 v_h^2~,~ 
m^2_{\eta_R^0}=m^2_{\eta_I^0}=m_2^2+\frac{1}{2}(\lambda_3+\lambda_4) v_h^2~,\label{mass}\\
m^2_{\phi_R}&= m_4^2+m_5^2+\gamma_5 v_h^2~,~m^2_{\phi_I}=m_4^2-m_5^2+\gamma_5 v_h^2~,\\
m^2_{\chi}&=m_3^2+\gamma_2 v_h^2~.
\end{align}
As we see from (\ref{mass}),
the tree-level mass of $\eta^0_R$ is the same as that of  $\eta^0_I$.
At the one-loop level, this degeneracy is lifted because 
the $\lambda_5$ term  is generated at this order:
\begin{eqnarray}
\lambda_5^{\rm eff} 
&=&
-\frac{\kappa^2}{64\pi^2}
\left[\frac{m^2_{\phi_I}}{m^2_{\phi_I}
-m_\chi^2}\ln \frac{m^2_{\phi_I}}{m_\chi^2}-
\frac{m^2_{\phi_R}}{m^2_{\phi_R}
-m_\chi^2}\ln \frac{m^2_{\phi_R}}{m_\chi^2}
\right]
\nonumber \\
&\sim&
-\frac{\kappa^2}{64\pi^2}
\frac{m_5^2}{m_{\phi_R}^2-m_{\chi}^2}
\left[
1-
\frac{m^2_{\chi}}{m^2_{\phi_R}-m_\chi^2}
\ln \frac{m^2_{\phi_R}}{m_\chi^2}
\right]~\mbox {for}~ m_5\ll m_{\phi_R}~.\label{l5}
\end{eqnarray}
In other words the origin of this correction is the one-loop self-energy diagram 
that can be embedded into 
 the two-loop diagram to generate  the neutrino mass (see Fig.~\ref{twoloop1}):
\begin{figure}
\includegraphics[width=8cm]{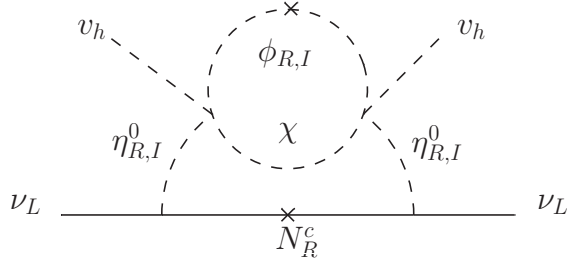}
\caption{\label{twoloop1}\footnotesize
The two-loop diagram that is responsible for the radiative generation 
of the neutrino mass.
The one-loop self-energy diagram inside of this two-loop diagram
is the origin of the mass difference between $m_{\eta_{R}^{0}}$ and 
$m_{\eta_{I}^{0}}$
as well as the effective coupling $\lambda^{\rm eff}_5$ in Eq.~(\ref{l5}).
}
\end{figure}
\begin{eqnarray}
({\cal M}_\nu)_{ij}  
&=&
\frac{Y^\nu_{ik}Y^\nu_{jk}}{16(4\pi)^4} \kappa^2 v_h^2
M_k (m_{\phi_I}^2 - m_{\phi_R}^2)
\int_0^1 dx_1 \int_0^{1-x_1}dx_2 \int_0^1dy
\left( \frac{y}{\beta-(x_1-x_1^2)M_k^2} \right)
\nonumber \\ &&
\times
\left[ \frac{x_1-x_1^2}{(1-y)\beta+y(x_1-x_1^2) m_{\eta^0}^2}
- \frac{1}{(1-y)M_k^2+y m_{\eta^0}^2}
\right]~,
\label{mnu0}
\end{eqnarray}
where $\beta \equiv m_{\phi_I}^2 +x_1(m_{\chi}^2 - m_{\phi_R}^2) +x_2(m_{\phi_I}^2 - m_{\phi_R}^2)$, and we have assumed that $
m_{\eta^{0}}=m_{\eta_{R}^{0}}\simeq 
m_{\eta_{I}^{0}}$.
Using $\lambda_5^{\rm eff}$ given in (\ref{l5}), 
the neutrino mass matrix can be approximated as
\begin{eqnarray}
({\cal M}_\nu)_{ij} 
&=& 
-\frac{\lambda_5^{\rm eff} v_h^2}{8 \pi^2}  
\sum_k 
\frac{Y^\nu_{ik} Y^\nu_{jk} M_k}{m_{\eta^{0}}^2 - M_k^2}
\left[ 1 - \frac{M_k^2}{m_{\eta^{0}}^2 - M_k^2}
 \ln \frac{m_{\eta^{0}}^2}{M_k^2} \right].
\label{mnu}
\end{eqnarray}

We see from  (\ref{mnu0}) that
the neutrino mass matrix ${\cal M}_\nu$ is proportional to 
$|Y^\nu \kappa|^2 m_5^2$
(because $ (m_{\phi_R}^2 - m_{\phi_I}^2)=2m^2_5$).
Therefore, only this combination 
for a given set of 
$m_{\chi}~,~m_{\phi_R}~,~m_{\eta_0}$ and $M_k$
can be fixed by the neutrino mass:
 $m_{\chi}$, $m_{\phi_R}$, $m_{\eta^0}$, $M_k
 \sim \mathcal{O}(10^2) ~\mathrm{GeV}$,
 for instance, implies that 
 $|Y^\nu \kappa| m_5 \sim \mathcal{O}(10^{-2}) ~\mathrm{GeV}$
  to obtain the  neutrino mass scale of $ \mathcal{O}(0.1) ~\mathrm{eV}$.
  With the same set of the parameter values we find that $\lambda^{\rm eff}_5
  \sim 10^{-4}$, where the smallness 
  $\lambda^{\rm eff}_5$ is a consequence of the radiative generation
  of this coupling.
  As we will see, the product $ |Y^\nu \kappa|$ enters into the semi-annihilation
  of DM particles that produces monochromatic neutrinos,
  while the upper bound of $|Y^\nu|$ follows from the $\mu \to e \gamma $ constraint.

\subsection{The stability of the scalar potential and
the perturbativity constraint}
If the parameters of the scalar potential 
$V=V_\lambda + V_m$ satisfy the following conditions, 
the potential is bounded from below and the DM stabilizing symmetry 
$Z_2 \times Z'_2$ remains unbroken at the tree-level:
\begin{eqnarray}
&&
m_1^2<0~,~~
m_2^2>0~,~~
m_3^2>0~,~~
m_4^2>0~,~~
|m_5^2|<m_4^2~,~ 
\nonumber \\ &&
\lambda_1 > 0~,~~
\lambda_2 > 0~,~~
\gamma_1 > 0~,~~
\gamma_4 > 0~,~ 
\nonumber \\ &&
\lambda_3 > -\frac{2}{3} \sqrt{\lambda_1 \lambda_2}~,~~
\lambda_3+\lambda_4 > -\frac{2}{3}\sqrt{\lambda_1 \lambda_2}~,~
 \nonumber \\ &&
\gamma_2 > -\frac{2}{3}\sqrt{\lambda_1 \gamma_1}~,~~
\gamma_3 > -\frac{2}{3}\sqrt{\lambda_2 \gamma_1}~,~~
\gamma_5 > -\frac{2}{3}\sqrt{\lambda_1 \gamma_4}~,~
 \nonumber \\ &&
\gamma_6 > -\frac{2}{3}\sqrt{\lambda_2 \gamma_4}~,~~
\gamma_7 > -\frac{2}{3}\sqrt{\gamma_1 \gamma_4}~,~ 
 \\ &&
|\kappa| < 
\lambda_1 + \lambda_2 + \gamma_1 + \gamma_4 
- \frac{2}{3} 
\left( 
\sqrt{\lambda_1 \lambda_2} +\sqrt{\lambda_1 \gamma_1}
+\sqrt{\lambda_1 \gamma_4} +\sqrt{\lambda_2 \gamma_1}
+\sqrt{\lambda_2 \gamma_4} +\sqrt{\gamma_1 \gamma_4}
\right)~.\nn
\label{stab}
\end{eqnarray}
We further assume that 
$|\lambda_i|$, $|\gamma_i|$, $|\kappa|$ $< 1 $
ensures the  perturbativeness of the model. Under these assumptions,
it is noted that 
the above stability conditions give $|\kappa|\lesssim 0.4$. 

\subsection{$\mu \rightarrow e \gamma$ constraint}
The strongest constraint on $Y^\nu$ comes from
$\mu\to e \gamma$
\footnote{The more detailed analysis of the lepton flavor violation such as the three body decays of lepton in the Ma model is discussed in Ref.\cite{Toma:2013zsa}.}, which  is given by \cite{Ma:2001mr,Adam:2013mnn}
\begin{eqnarray}
&&B(\mu\rightarrow e\gamma)={3\alpha\over 64
\pi(G_Fm_{\eta^\pm}^2)^2}
\left| ~\sum_k Y^\nu_{\mu k}Y^\nu_{ek}F_2
\left({M_k^2\over m_{\eta^\pm}^2}\right)
\right|^2 \lsim 5.7\times10^{-13}~,\label{mug}\\
&&F_2(x)={1\over 6(1-x)^4}(1-6x+3x^2+2x^3-6x^2\ln x)~.\nonumber
\end{eqnarray}  
A similar, but slightly weaker bound for $\tau \to \mu (e) \gamma$ given
in \cite{Hayasaka:2010et} has to be satisfied, too.
Since $F_2(x) \sim 1/3x$ for $x\gg 1$, while  
$1/12 < F_2(x) < 1/6 $ for $0< x <1$, the constraint 
can be readily  satisfied if
$M_{k}\gg m_{\eta^\pm}$ or $M_{k} < m_{\eta^\pm}$.
If we assume that $M_k\sim m_{\eta^\pm} \sim \mathcal{O}(10^2) ~\mbox{GeV}$
in (\ref{mug}),
the constraint  (\ref{mug}) becomes
$B(\mu\rightarrow e\gamma)\simeq
10^{-4}\times |\sum_k Y^\nu_{\mu k}Y^\nu_{ek}|^2
  \lsim 5.7\times10^{-13}$.
Therefore, $|Y^\nu_{e k}Y^\nu_{\mu k}|^2\lsim \mathcal{O}(10^{-8})$
can satisfy the constraint.

\subsection{The $T$ parameter constraint}
Of the $S,T$ and $U$ parameters 
from the electroweak precision measurements the  $T$ parameter constraint
is the severest for the present model
\cite{Barbieri:2006dq,Beringer:1900zz,Ciuchini:2013pca},
\be
\Delta T &\simeq & 1.08
\left(\frac{m_{\eta^\pm}-m_{\eta^0_R}}{v_h}\right)
\left(\frac{m_{\eta^\pm}-m_{\eta^0_I}}{v_h}\right)
=0.10\pm 0.08 \label{cnd4}
\ee
for $m_h=125.6\pm0.3$ GeV.
Therefore,  
$|m_{\eta^\pm}-m_{\eta^0_R}|~,~
|m_{\eta^\pm}-m_{\eta^0_I}| \lsim 90~\mbox{GeV}$ 
is sufficient to meet the requirement.

\begin{figure}
\centering
\begin{minipage}{0.3\textwidth}
	\includegraphics[width=40mm]{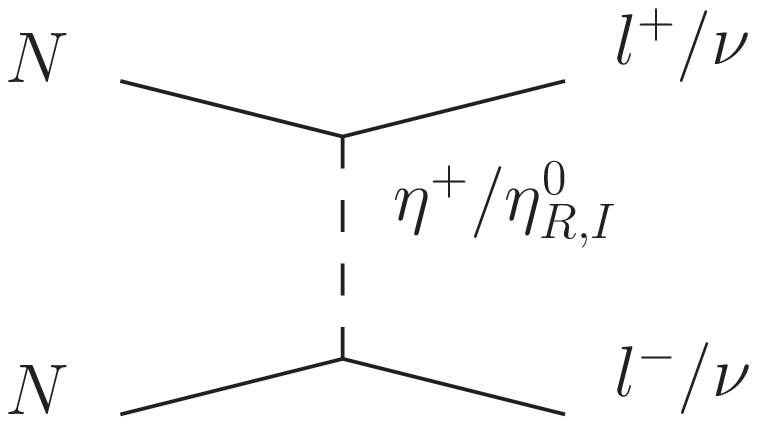}
\end{minipage}
\begin{minipage}{0.3\textwidth}
	\includegraphics[width=40mm]{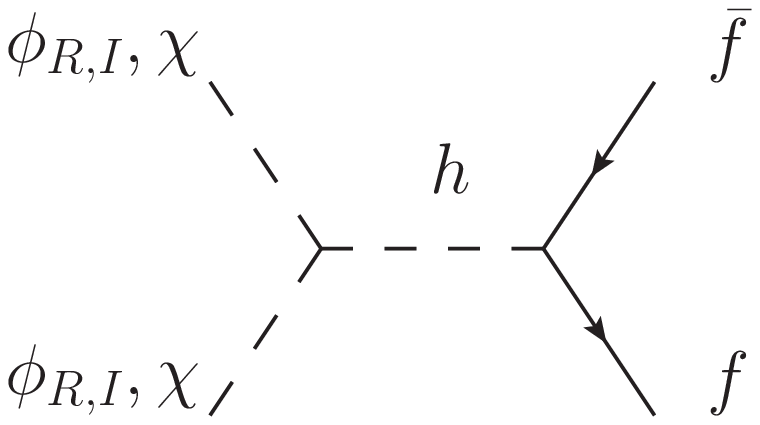}
\end{minipage}
\begin{minipage}{0.3\textwidth}
	\includegraphics[width=40mm]{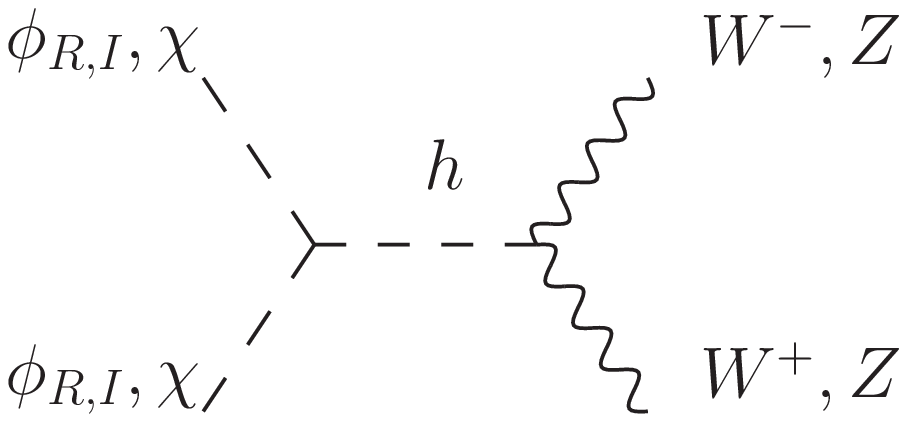}
\end{minipage}
\begin{minipage}{0.9\textwidth}
	\includegraphics[width=120mm]{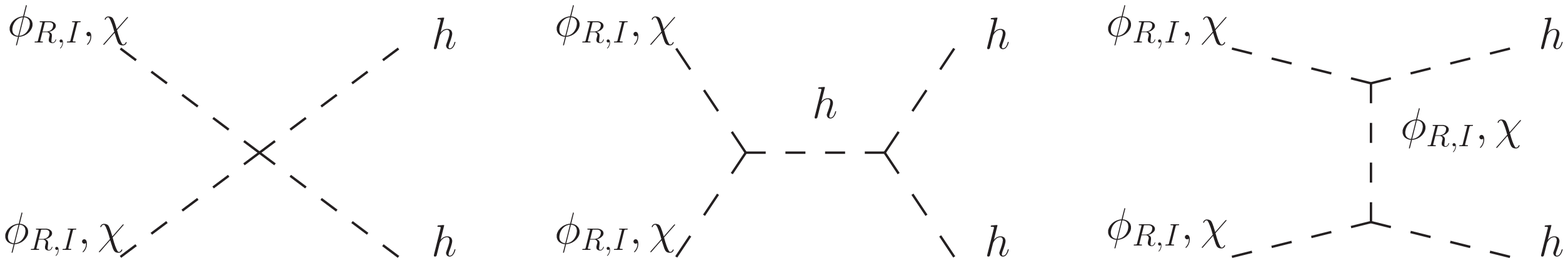}
\end{minipage}
	\caption{The diagrams for the standard annihilation processes. }
	\label{fig:stand}
\end{figure}
\begin{figure}
\centering
	\includegraphics[width=120mm]{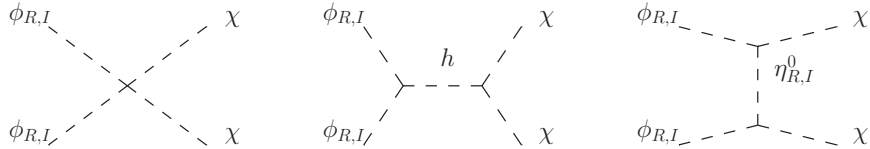}
	\caption{The diagrams for the DM conversion processes.}
	\label{fig:DM_conv}
\end{figure}
\begin{figure}
\centering
	\includegraphics[width=40mm]{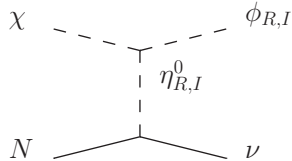}
	\caption{The diagrams for the semi-annihilation process.}
	\label{fig:semi}
\end{figure}

\section{Multicomponent dark matter system}
In this model there are three types of dark matter candidates 
$N= N^c_{R1}$ (the lightest among $N^c_{Rk}$'s) or $\eta^0_R$ (or $\eta^0_I$) with $(Z_2, Z'_2)=(-,+)$, 
$\chi$ with $(Z_2, Z'_2)=(+,-)$ and $\phi_R$ (or  $\phi_I$) with $(Z_2, Z'_2)=(-,-)$. 
For $(Z_2, Z'_2)=(-,+)$ there are two candidates, and in the following discussions we assume that  $N$ is a DM  candidate
\footnote{
The other possibility, $\eta_R^0$-DM, is discussed in \cite{Aoki:2013gzs}.
}.
Therefore, our system consists of three DM particles, $N ,~ \phi_R,~ \chi$. 
Consequently,
there are three types of DM annihilation process \cite{Aoki:2012ub} (Figures \ref{fig:stand}-\ref{fig:semi}); 
\begin{align}
\mathrm{Standard~annihilation:}&~
NN \rightarrow XX',~~\phi_R \phi_R \rightarrow XX',~~\chi\chi \rightarrow XX', ~~(X,X' \mathrm{:SM~particles}),~\\
\mathrm{DM~conversion:}&~\phi_R \phi_R \rightarrow \chi\chi,~\\
\mathrm{Semi}\rm{\mathchar`-annihilation:}&~
N \phi_R \rightarrow \chi \nu,~~
\chi N  \rightarrow \phi_R \nu,~~
\phi_R \chi \rightarrow N \nu,~
\end{align}
\noindent
where we assume $m_{\phi_R} > m_\chi$.
Moreover, since the mass difference between $\phi_R$ and $\phi_I$ is 
controlled by the lepton-number breaking mass $m_5$, which is assumed 
to be much smaller than $m_{\phi_R}$ so that $m_{\phi_R}$ and $m_{\phi_I}$ are practically degenerate, the contribution of $\phi_I$ to the annihilation processes during the decoupling of DMs is non-negligible. 
The annihilation processes of $\phi_I$ are
$\phi_I \phi_I \rightarrow XX'$ (standard annihilation),~
$\phi_I \phi_I \rightarrow \phi_R \phi_R$ and $\phi_I \phi_I \rightarrow \chi\chi$ (DM conversion),
$N \phi_I \rightarrow \chi \nu,~
\chi N  \rightarrow \phi_I \nu$ and
$\phi_I \chi \rightarrow N \nu~$ (semi-annihilation),
where we have assumed that the decay of $\phi_I \rightarrow N \chi$ is kinematically forbidden. 
There is a conversion between $\phi_R$ and $\phi_I$, and its
 main process is shown in Fig.~\ref{fig:phi_IR_loop}. 
This process is loop suppressed, and the cross section 
\begin{align}
\sigma_{\phi_I X \rightarrow \phi_R X'} |v| 
\sim
10^{-14}  \times
\left( \frac{ v_h }{ 246 ~\mathrm{GeV} } \right)^4
\left( \frac{ m_{\phi}}{100 ~\mathrm{GeV} } \right)^2
\left( \frac{ (100 ~\mathrm{GeV})^6 }{ m_{\eta^0_R}^2 m_{\eta^0_I }^2 m_{\chi}^2 } \right)
~\mathrm{GeV}^{-2} ~
\end{align}
would be roughly 2 orders of magnitude  smaller than that of tree-level processes.
The reaction rate of this process is $\langle \sigma_{\phi_I X \rightarrow \phi_R X'} |v| \rangle n_{\phi_I} n_X$ 
which is roughly $n_{\phi_I}^{-1} \sim \exp(m_{\phi_I}/T)$ times larger than the standard annihilation. 
Thus, during the decoupling of DMs, the reaction between $\phi_I$ and $\phi_R$ can reach chemical equilibrium, implying that
\begin{figure}
\centering
	\includegraphics[width=45mm]{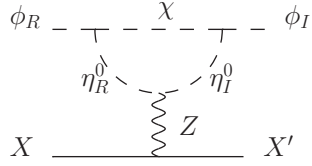}
	\caption{Conversion process between $\phi_I$ and $\phi_R$. Here $X$ and $X'$ are SM particles.}
	\label{fig:phi_IR_loop}
\end{figure} 
 we can use a similar method as \cite{Griest:1990kh} and sum up the number densities of particles having the same $Z_2 \times Z'_2$ parities. 
The Boltzmann equations of their number densities $n_N,~n_\phi \equiv n_{\phi_I}+n_{\phi_R},~n_\chi$ are given by
\begin{align}
\dot{n}_N + 3Hn_N
=&
-\Bigl\{ 
\langle \sigma_{NN \rightarrow XX'} |v| \rangle (n_N^2- \bar{n}_N^2 )
\nonumber \\ &
+ \langle \sigma_{N \phi \rightarrow \chi \nu} |v| \rangle
(n_N n_\phi - \bar{n}_N \bar{n}_\phi \frac{ n_\chi }{ \bar{n}_\chi })
+ \langle \sigma_{N \chi \rightarrow \phi \nu} |v| \rangle
(n_N n_\chi - \bar{n}_N \bar{n}_\chi \frac{ n_\phi }{ \bar{n}_\phi })
\nonumber \\ &
- \langle \sigma_{\phi \chi \rightarrow N \nu} |v| \rangle
(n_\phi n_\chi - \bar{n}_\phi \bar{n}_\chi \frac{ n_N }{ \bar{n}_N })
\Bigr\}~,
\label{eq:boltz1}
\\
\dot{n}_\phi + 3Hn_\phi
=&
-\Bigl\{ 
\langle \sigma_{\phi \phi \rightarrow XX'} |v| \rangle
(n_\phi^2- \bar{n}_\phi^2 )
+ \langle \sigma_{\phi\phi \rightarrow \chi\chi} |v| \rangle
(n_\phi^2- \bar{n}_\phi^2 \frac{ n_\chi^2 }{ \bar{n}_\chi^2 } )
\nonumber \\ &
+ \langle \sigma_{N \phi \rightarrow \chi \nu} |v| \rangle
(n_N n_\phi - \bar{n}_N \bar{n}_\phi \frac{ n_\chi }{ \bar{n}_\chi })
- \langle \sigma_{N \chi \rightarrow \phi \nu} |v| \rangle
(n_N n_\chi - \bar{n}_N \bar{n}_\chi \frac{ n_\phi }{ \bar{n}_\phi })
\nonumber \\ &
+ \langle \sigma_{\phi \chi \rightarrow N \nu} |v| \rangle
(n_\phi n_\chi - \bar{n}_\phi \bar{n}_\chi \frac{ n_N }{ \bar{n}_N })
\Bigr\}~,
\label{eq:boltz2}
\\
\dot{n}_\chi + 3Hn_\chi
=&
-\Bigl\{ 
\langle \sigma_{\chi \chi \rightarrow XX'} |v| \rangle
(n_\chi^2- \bar{n}_\chi^2 )
- \langle \sigma_{\phi\phi \rightarrow \chi\chi} |v| \rangle
(n_\phi^2- \bar{n}_\phi^2 \frac{ n_\chi^2 }{ \bar{n}_\chi^2 } )
\nonumber \\ &
- \langle \sigma_{N \phi \rightarrow \chi \nu} |v| \rangle
(n_N n_\phi - \bar{n}_N \bar{n}_\phi \frac{ n_\chi }{ \bar{n}_\chi })
+ \langle \sigma_{N \chi \rightarrow \phi \nu} |v| \rangle
(n_N n_\chi - \bar{n}_N \bar{n}_\chi \frac{ n_\phi }{ \bar{n}_\phi })
\nonumber \\ &
+ \langle \sigma_{\phi \chi \rightarrow N \nu} |v| \rangle
(n_\phi n_\chi - \bar{n}_\phi \bar{n}_\chi \frac{ n_N }{ \bar{n}_N })
\Bigr\}~,
\label{eq:boltz3}
\end{align}
where $H$ is the Hubble parameter. We have made approximations given by
\begin{align}
\langle \sigma_{\phi \phi \rightarrow XX'} |v| \rangle
&=
\langle \sigma_{\phi_I \phi_I \rightarrow XX'} |v| \rangle
\frac{\bar{n}_{\phi_I}^2}{\bar{n}_{\phi}^2}
+\langle \sigma_{\phi_R \phi_R \rightarrow XX'} |v| \rangle
\frac{\bar{n}_{\phi_R}^2}{\bar{n}_{\phi}^2}~,
\\
\langle \sigma_{\phi\phi \rightarrow \chi\chi} |v| \rangle
&=
\langle \sigma_{\phi_I \phi_I \rightarrow \chi\chi} |v| \rangle
\frac{\bar{n}_{\phi_I}^2}{\bar{n}_{\phi}^2}
+\langle \sigma_{\phi_R \phi_R \rightarrow \chi\chi} |v| \rangle
\frac{\bar{n}_{\phi_R}^2}{\bar{n}_{\phi}^2}~,
\\
\langle \sigma_{N \phi \rightarrow \chi \nu} |v| \rangle
&=
\langle \sigma_{N \phi_I \rightarrow \chi \nu} |v| \rangle
\frac{\bar{n}_{\phi_I}}{\bar{n}_{\phi}}
+\langle \sigma_{N \phi_R \rightarrow \chi \nu} |v| \rangle
\frac{\bar{n}_{\phi_R}}{\bar{n}_{\phi}}~,
\\
\langle \sigma_{N \chi \rightarrow \phi \nu} |v| \rangle
&=
\langle \sigma_{N \chi \rightarrow \phi_I \nu} |v| \rangle
+\langle \sigma_{N \chi \rightarrow \phi_R \nu} |v| \rangle~,
\\
\langle \sigma_{\phi \chi \rightarrow N \nu} |v| \rangle
&=
\langle \sigma_{\phi_R \chi \rightarrow N \nu} |v| \rangle
\frac{\bar{n}_{\phi_R}}{\bar{n}_{\phi}}
+\langle \sigma_{\phi_I \chi \rightarrow N \nu} |v| \rangle
\frac{\bar{n}_{\phi_I}}{\bar{n}_{\phi}}~.
\end{align}
As usual we rewrite (\ref{eq:boltz1}), (\ref{eq:boltz2}) and  (\ref{eq:boltz3}) 
for $Y_i \equiv n_i/s$, where $s$ is the entropy density. 
To this end, we introduce the reaction rates
\begin{align}
\Gamma_{i i \rightarrow XX'} 
&= 
\langle \sigma_{i i \rightarrow XX'} |v| \rangle \bar{n}_i ~,
\\
\Gamma^{(i)}_{j j \rightarrow k k} 
&= 
\langle \sigma_{j j \rightarrow k k} |v| \rangle  
\frac{\bar{n}_j^2}{\bar{n}_i}
~~(\mathrm{for}~i=j~\mathrm{or}~k, m_j>m_k)~,
\\
\Gamma^{(i)}_{j k \rightarrow l X} 
&= 
\langle \sigma_{j k \rightarrow l X} |v| \rangle 
\frac{ \bar{n}_j \bar{n}_k}{\bar{n}_i}
~~(\mathrm{for}~i=j,~k~\mathrm{or}~l)
~.
\end{align}
For $i=j=\phi$, $k=\chi$, for instance, the DM conversion rate is 
$\Gamma^{(\phi)}_{\phi\phi \rightarrow \chi\chi}$, 
while is $\Gamma^{(\chi)}_{\phi\phi \rightarrow \chi\chi}$ for 
$i=k=\chi$, $j=\phi$. 
The ratio between $\Gamma^{(\chi)}_{\phi\phi \rightarrow \chi\chi}$ and 
$\Gamma^{(\phi)}_{\phi\phi \rightarrow \chi\chi}$ is given by the factor 
$\bar{n}_\phi /\bar{n}_\chi$, which is small because $m_{\phi} > m_{\chi}$. 
Similarly, the ratio of the semi-annihilation process and the standard annihilation for the DM $i$ is proportional to 
${ \bar{n}_j \bar{n}_k}/{\bar{n}_i^2}\sim\exp(-(m_j+m_k-2m_i)/T)= \exp(\pm(m_j-m_k)/T)$
for $i=j$ or $k$.
If the $\langle \sigma |v| \rangle $'s are the same order of magnitude,
this factor implies a larger rate in the Boltzmann  equations for the heavier DM and a smaller rate for lighter DM.
In the case $i=l$, the factor is ${ \bar{n}_j \bar{n}_k}/{\bar{n}_i^2}\sim\exp(-(m_j+m_k-2m_l)/T)$ and it can be enhanced 
when $(m_j+m_k)/2 < m_l$.
Using these reaction rates we find 
\begin{align}
\frac{x}{\bar{Y}_N}
\frac{dY_N}{dx}=&
- \frac{\Gamma_{NN \rightarrow XX'}}{H(x)}
\left(\frac{Y_N^2}{\bar{Y}_N^2}-1 \right)
- \frac{\Gamma^{(N)}_{N\phi \rightarrow \chi \nu}}{H(x)}
\left( \frac{Y_N Y_\phi}{\bar{Y}_N \bar{Y}_\phi}- \frac{Y_\chi}{\bar{Y}_\chi} \right)
\nonumber \\ &
- \frac{\Gamma^{(N)}_{N\chi \rightarrow \phi \nu}}{H(x)}
\left( \frac{Y_N Y_\chi}{\bar{Y}_N \bar{Y}_\chi}- \frac{Y_\phi}{\bar{Y}_\phi} \right)
+ \frac{\Gamma^{(N)}_{\phi \chi \rightarrow N \nu}}{H(x)}
\left( \frac{Y_\phi Y_\chi}{\bar{Y}_\phi \bar{Y}_\chi}- \frac{Y_N}{\bar{Y}_N} \right)
~,\\
\frac{x}{\bar{Y}_\phi}
\frac{dY_\phi}{dx}=&
-\frac{\Gamma_{\phi \phi \rightarrow XX'}}{H(x)}
\left(\frac{Y_\phi^2}{\bar{Y}_\phi^2}-1 \right)
- \frac{\Gamma^{(\phi)}_{\phi\phi \rightarrow \chi \chi}}{H(x)}
\left( \frac{Y_\phi^2}{\bar{Y}_\phi^2}- \frac{Y_\chi^2}{\bar{Y}_\chi^2} \right)
- \frac{\Gamma^{(\phi)}_{N\phi \rightarrow \chi \nu}}{H(x)}
\left( \frac{Y_N Y_\phi}{\bar{Y}_N \bar{Y}_\phi}- \frac{Y_\chi}{\bar{Y}_\chi} \right)
\nonumber \\ &
+ \frac{\Gamma^{(\phi)}_{N\chi \rightarrow \phi \nu}}{H(x)}
\left( \frac{Y_N Y_\chi}{\bar{Y}_N \bar{Y}_\chi}- \frac{Y_\phi}{\bar{Y}_\phi} \right)
- \frac{\Gamma^{(\phi)}_{\phi \chi \rightarrow N \nu}}{H(x)}
\left( \frac{Y_\phi Y_\chi}{\bar{Y}_\phi \bar{Y}_\chi}- \frac{Y_N}{\bar{Y}_N} \right)
~,\\
\frac{x}{\bar{Y}_\chi}
\frac{dY_\chi}{dx}=&
-\frac{\Gamma_{\chi \chi \rightarrow XX'}}{H(x)}
\left(\frac{Y_\chi^2}{\bar{Y}_\chi^2}-1 \right)
+ \frac{\Gamma^{(\chi)}_{\phi\phi \rightarrow \chi \chi}}{H(x)}
\left( \frac{Y_\phi^2}{\bar{Y}_\phi^2}- \frac{Y_\chi^2}{\bar{Y}_\chi^2} \right)
+ \frac{\Gamma^{(\chi)}_{N\phi \rightarrow \chi \nu}}{H(x)}
\left( \frac{Y_N Y_\phi}{\bar{Y}_N \bar{Y}_\phi}- \frac{Y_\chi}{\bar{Y}_\chi} \right)
\nonumber \\ &
- \frac{\Gamma^{(\chi)}_{N\chi \rightarrow \phi \nu}}{H(x)}
\left( \frac{Y_N Y_\chi}{\bar{Y}_N \bar{Y}_\chi}- \frac{Y_\phi}{\bar{Y}_\phi} \right)
- \frac{\Gamma^{(\chi)}_{\phi \chi \rightarrow N \nu}}{H(x)}
\left( \frac{Y_\phi Y_\chi}{\bar{Y}_\phi \bar{Y}_\chi}- \frac{Y_N}{\bar{Y}_N} \right)
~,
\end{align}
where $x=\mu/T$, $\mu=(M_1+m_{\phi_R}+m_{\chi})/3,$ $H(x)=1.67g_*^{1/2} \mu^2 /m_{pl}/ x^{2} $ and $g_*$ is the effective degrees of freedom of the massless particle in the Universe.

In the original Ma model \cite{Ma:2006km}, the relic density of $N$ tends to be larger than the observational value \cite{{Kubo:2006yx}}. 
The additional contributions coming from the semi-annihilation can enhance the annihilation rate for $N$ so that the $N$ DM contribution to $\Omega h^2$ can be suppressed. 
In this way the tension between the constraint from lepton flavor violation and the cosmological observation of $\Omega h^2$ may become mild in the present model. 

There are many mass parameters in the model, on which the relic abundance of DM depends. As a benchmark run, we vary $m_\chi$ from 135 GeV to 300 GeV with the fixed right-handed neutrino masses
$M_1=300$ GeV and $M_2=M_3=1$ TeV,
while the other masses are varied with a fixed mass deference relative to $m_\chi$ i.e.  
$m_{\eta^0_R}=m_\chi + m_{\phi_R} - 10~ \mathrm{GeV},~
m_{\phi_I}=m_{\phi_R}+10 ~\mathrm{GeV}$, and 
$m_{\phi_R}=m_{\chi}+50 ~\mathrm{GeV}$.  
Moreover, for simplicity, we use the common size of the scalar couplings, i.e. 
$ \gamma \equiv \gamma_2=\gamma_5=\gamma_7$ . 
The mass differences are chosen so that no resonance appears in the s-channel of the semi-annihilation, i.e. $m_{\eta^0_{R,I}} < m_{\phi_{R,I}}+m_{\chi}$.
Fig.~\ref{fig:om_mchi} shows the $m_\chi$ dependence of the individual relic densities for $\gamma=0.1$, where the input parameters are summarized in Table \ref{tab:param_fig6}. 
When the scalar particles involved in  the semi-annihilation 
are lighter than $N$, the semi-annihilation tends to decrease
 the relic density of  the $N$ DM (blue, dashed line). 
The total relic density of DM can be made consistent with
the observed value $\Omega h^2 \sim 0.12$ \cite{Hinshaw:2012aka,Ade:2013zuv} by varying the size of the scalar couplings. 
Fig.~\ref{fig:gamma_mass} is a contour plot for the $m_\chi$-$\gamma$ plane. 
The scalar coupling $\gamma$ that is consistent with $\Omega h^2 \sim 0.12$ increases drastically 
at $m_{\chi}\sim220$ GeV because the relic density of the $N$ DM $\Omega_N h^2$ becomes close to 0.12 at $m_\chi \sim220$ 
GeV (as one can see from Fig.~\ref{fig:om_mchi}), so that $\Omega_\phi h^2$ and $\Omega_\chi h^2$ should be drastically suppressed.

\begin{table}
\centering
\caption{Parameter set for the calculation of Fig.\ref{fig:om_mchi}.}
\label{tab:param_fig6}
\begin{tabular}{|c|c|c|c|c|c|c|c|c|} \hline
 $M_1$                    & $M_{2},M_3$        & $m_{\eta^+}$     &$m_{\eta^0_R}$ &$m_{\phi_I}$       &$m_{\phi_R}$      &$\gamma$            & $\kappa$              & $Y^\nu$               
 \\ \hline 
  300 GeV & 1 TeV & $m_{\eta^0_R} -10 ~\mathrm{GeV}$ & $ m_\chi +m_{\phi_R} - 10~ \mathrm{GeV}$ & $m_{\chi}+60 ~\mathrm{GeV}$ & $m_{\chi}+50 ~\mathrm{GeV}$ & $0.1$ & $0.4$ & $0.01$ \\ \hline
\end{tabular}
\end{table}
\begin{figure}
\includegraphics[width=8cm]{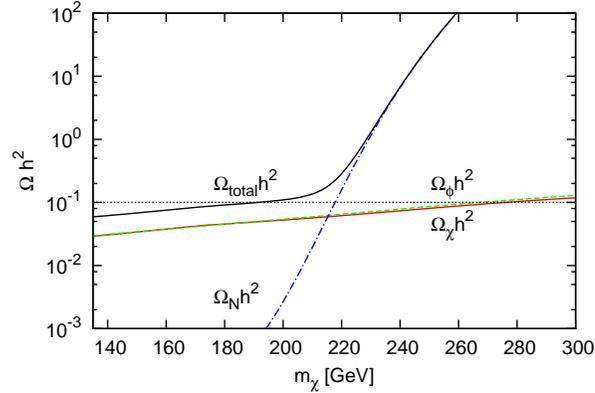}
\caption{The $m_\chi$ dependence of the relic density 
$\Omega_{\chi} h^2$ (red solid line), $\Omega_{\phi} h^2$ (green dashed line), $\Omega_{N} h^2$ (blue dot-dashed line) and $\Omega_{\rm total} h^2$ (black solid line). The fixed parameters are shown in Table \ref{tab:param_fig6}. }
\label{fig:om_mchi}
\end{figure}

\begin{figure}
\centering
\vspace{8mm}
	\includegraphics[width=8cm]{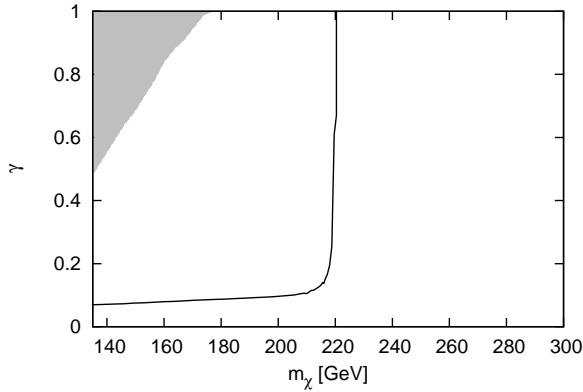}
	\caption{Contour plot for the total relic density $\Omega_{\rm total} h^2 \sim 0.12$. The gray region is excluded by the constraint of vacuum stability. The threshold value of $m_\chi$ depends in particular on $M_1$. For the parameters given in Table \ref{tab:param_fig6}, except for $M_1=500$ GeV, we find $m_\chi \sim 380$ GeV, for instance.    }
	\label{fig:gamma_mass}
\end{figure}

\subsection{Direct detection}
The current upper bound for the DM-nucleon cross section  is  estimated assuming the  one-component DM scenario and current upper bound and future sensitivity are given in Refs. \cite{ Aprile:2012nq,Aprile:2012zx,Akerib:2013tjd}.
Because the collision rate is roughly proportional to $\sigma n_{_\mathrm{DM}}$, the upper bound for the event rate can be translated to the constraint on the detection rate in the multicomponent DM scenario. 
The effective cross section of the nucleon corresponding to the cross section of the nucleon in the one-component DM scenario is given by 
\begin{eqnarray}
\sigma^{\mathrm{eff}}_i
=
\sigma_i 
\left( \frac{ \Omega_i h^2 }{ \Omega_{\mathrm{total}} h^2 } \right)
~. \label{ddsigma}
\end{eqnarray}
In our model, only $\phi_R$ and $\chi$ DM scatter with the nucleus, and  
the right-handed neutrino DM $N$ does not interact with nucleus at tree level. 
So we can  neglect the  $N$ contribution at the lowest order
in perturbation theory.  The cross sections of $\phi_R$ and $\chi$ are given by \cite{Barbieri:2006dq}
\begin{eqnarray}
\sigma_{\phi_R}
&=&
\frac{1}{\pi}
\left(  \frac{ (\gamma_5/2) \hat{f} m_N }{ m_{\phi_R} m_h }  \right)^2
\left(  \frac{ m_N m_{\phi_R} }{ m_N + m_{\phi_R} }  \right)^2 ~,
 \\
 \sigma_{\chi}
&=&
\frac{1}{\pi}
\left(  \frac{ \gamma_2 \hat{f} m_N }{ m_{\chi} m_h }  \right)^2
\left(  \frac{ m_N m_{\chi} }{ m_N + m_{\chi} }  \right)^2 ~,
\end{eqnarray}
where $\hat{f} \sim 0.3$ is the usual nucleonic matrix element \cite{Ellis:2000ds}, and $m_N$ is the nucleon mass. 
Fig.~\ref{DDreson10} shows the relation between $m_\chi$ and 
the sum of the  effective cross sections given in (\ref{ddsigma}).
The black line corresponds to the parameter space
(the black line in Fig.~\ref{fig:gamma_mass})  consistent with  the 
cosmological observation of the DM relic abundance. 
Although as we see from
Fig.~\ref{fig:gamma_mass}, the scalar coupling $\gamma$ has to become large at $m_\chi \sim 220 ~\mathrm{GeV}$,
such that the cross sections off the nucleon,
$\sigma_{\chi}$ and $\sigma_{\phi_R}$, 
become large, 
$\sigma_{\phi_R}^{\rm eff}+\sigma_{\chi}^{\rm eff}$ 
does not change very much at 
$m_\chi \sim 220 ~\mathrm{GeV}$, because 
$\Omega_{\phi_R}$ and $\Omega_{\chi}$ both 
become small. We also show the result for $M_{1}=500$ GeV (the red line)
in Fig.~\ref{DDreson10}, where the other parameters are taken as the same as in the case with 
$M_{1}=300$ GeV.

\begin{figure}
\centering
\includegraphics[width=9cm]{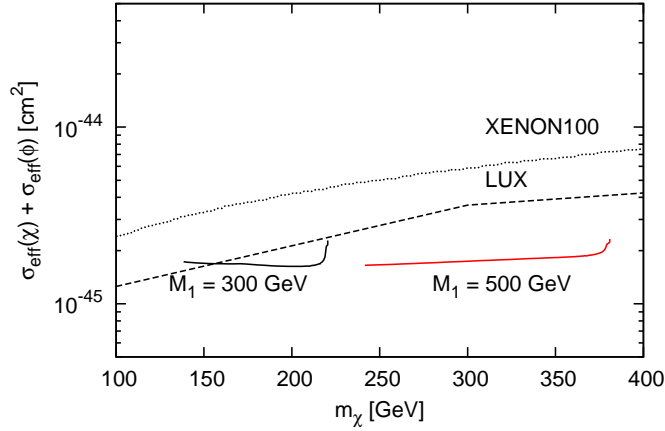}
\caption{\label{DDreson10} 
\footnotesize 
The relation between the $\chi$ DM mass 
$m_{\chi}$  and 
the sum of the  effective cross sections given in (\ref{ddsigma}).
The black (red) line shows the result for $M_1=300$ (500) GeV.
 The black dotted and dashed lines show the upper limit of the spin independent cross section 
 off the nucleon given by XENON100
 \cite{Aprile:2012nq} and LUX \cite{Akerib:2013tjd}, respectively.}
\end{figure}

\subsection{Indirect detection} 
For indirect detections of DM 
 the SM particles produced by the annihilation of DM are searched.
Because the semi-annihilation produces a SM particle, this process can 
serve for an indirect detection. 
In our model, especially, the SM particle from the semi-annihilation process as
shown in Fig.~\ref{fig:semi} is neutrino which has a monochromatic energy spectrum \cite{Aoki:2012ub}. 
Therefore,  we consider below the neutrino flux from the Sun 
\cite{Silk:1985ax,
 Krauss:1985aaa, 
 Freese:1985qw,
Gaisser:1986ha, 
Griest:1986yu, 
Ritz:1987mh, 
Kamionkowski:1991nj,
 Kamionkowski:1994dp} 
as a possibility to detect the semi-annihilation process of DMs.

The DM particles are captured in the Sun losing their 
kinematic energy through scattering with the nucleus. 
Then captured DM particles annihilate each other.
The time dependence of the  number of DM $n_i$ in the Sun is given by
\begin{eqnarray}
\dot{n}_i
&=&
C_i
 - C_A(ii \rightarrow \mathrm{SM})n_i^2
 - \sum_{m_i>m_j}C_A(ii \rightarrow jj)n_i^2
 - C_A(ij \rightarrow k \nu)n_i n_j~,
 \label{Boltzsun}
\end{eqnarray}
where $C_i$ is the capture rate in the Sun, and $C_A$'s are the annihilation
rates in the Sun\cite{Kamionkowski:1991nj,Kamionkowski:1994dp,Griest:1986yu}:
\begin{eqnarray}
C_\chi &\sim &
 1.4\times 10^{20} f(m_{\chi})\left( \frac{\hat{f}}{0.3} \right)^2 
\left( \frac{\gamma_{2}}{0.1} \right)^2 
\left( \frac{200~\mathrm{GeV}}{m_{\chi}} \right)^2 
\left( \frac{\Omega_{\chi} h^2}{ \Omega_{ \mathrm{total} } } \right)^2 
~,\\
C_{\phi_R} &\sim &
 1.4\times 10^{20} f(m_{\phi_R})\left( \frac{\hat{f}}{0.3} \right)^2 
\left( \frac{\gamma_{5}}{0.2} \right)^2 
\left( \frac{200~\mathrm{GeV}}{m_{\phi_R}} \right)^2 
\left( \frac{\Omega_{\phi_R} h^2}{ \Omega_{ \mathrm{total} } } \right)^2 
~,\\
C_{N} &= & 0~,
\\
C_A(ij \rightarrow \bullet)
&=&
\frac{ \langle \sigma(ij \rightarrow \bullet)|v| \rangle }
{ V_{ij} }
~,~~~
V_{ij} = 5.7 \times 10^{27}
\left( \frac{100~\mathrm{GeV}}{\mu_{ij}} \right)^{3/2} \mathrm{cm}^3 ~.
\end{eqnarray}
Here $f(m_{i})$ depends on the form factor of the nucleus, elemental abundance, kinematic suppression of the capture rate, etc.,
varying ${\mathcal O}(0.01 - 1)$ depending on the DM mass \cite{Kamionkowski:1991nj,Kamionkowski:1994dp}.
$V_{ij}$ is an effective volume of the Sun with $\mu_{ij}=2m_i m_j / (m_i+m_j)$
 in the nonrelativistic limit. 
We neglect the DM production processes in Eq.(\ref{Boltzsun})  like $jj \rightarrow ii$ and $jk \rightarrow iX$ because the kinetic energy of the produced particle $i$ is much larger than  that corresponding to
the escape velocity from  the Sun, i.e. $\sim 10^3$ km/s
\cite{Griest:1986yu,Agrawal:2008xz}. 
Consequently,  the number of the right-hand neutrino DM cannot increase, 
and hence  $\phi\chi \rightarrow N \nu$ is   the only neutrino  production process,
where
its reaction rate is given by 
$\Gamma(\nu) = C_A(\phi \chi \rightarrow N \nu)n_\phi n_\chi $
\footnote{There are also neutrinos having a continuous energy spectrum from the decay of standard model particles, $W^+$ or $b$ for instance, produced by standard annihilation of scalar DMs. The upper bounds for the production rates of the standard model particles are given in\cite{IceCube:2011aj,Aartsen:2012kia,Agrawal:2008xz}. }.

The monochromatic neutrino flux on the Earth is roughly given by 
$\Gamma_{\mathrm inc} = \Gamma /4\pi R_\odot^2$,
where $R_\odot$ stands for the distance to the Sun. 
Fig.~\ref{nufluxreson} shows the $m_{\chi}$ dependence 
of the neutrino flux for the same parameter  space (black line) as in Fig.~\ref{fig:om_mchi}.
As we can see from Fig.~\ref{fig:semi}
a resonance effect for
the s-channel annihilation process  can be achieved if 
$m_{\eta^0_R} \simeq m_{\phi_R}+m_\chi$.
Obviously, the smaller the mass difference 
$m_{\eta^0_R} -(m_{\phi_R}+m_\chi )$ is, 
the larger is  the semi-annihilation cross section
 and hence the neutrino flux. 
In Fig.~\ref{nufluxreson} four different  values are used:
$\left(m_{\eta^0_R} -(m_{\phi_R}+m_\chi)~,~M_1\right)=$ 
(10 GeV, 300 GeV) (black curve), (1 GeV, 300 GeV) (magenta curve), (10 GeV, 500 GeV) (red curve), and (1 GeV, 500 GeV) (blue curve), respectively.
In the case that $\Omega_{N}$
dominates (so that $\Omega_{\phi}$ and $\Omega_{\chi}$ are
small) the capture rates of the $\phi$ and  $\chi$ DMs become small
(see (\ref{Boltzsun})).
This is why the neutrino flux decreases after a certain value of $m_{\chi}$. 

The upper limits on the defused neutrino flux from the Sun are given 
 by the IceCube experiment \cite{Aartsen:2012kia}.
The upper limit on the neutrino flux produced  by 
 the annihilation of the DMs 
of $250$ GeV into $W^+W^-$,
for instance, 
is $9.72 \times 10^{10}$ $\mathrm{ km}^{-2}\mathrm{y}^{-1}$ \cite{Aartsen:2012kia}. 
We can see from Fig.~\ref{nufluxreson} 
that, unfortunately,  this limit is at least $10^3$ times larger than the monochromatic neutrino flux produced by the semi-annihilation of the $\phi$ and $\chi$ DMs. 
Note however, that the energy spectrum of the neutrino flux produced by the $W$ decay is different from the monochromatic neutrino.
With  an increasing resolution of energy and angle
 the  chance for the observation of the semi-annihilation
 and hence of a multicomponent nature of DM can increase.

\begin{figure}
\centering
	\includegraphics[width=8cm]{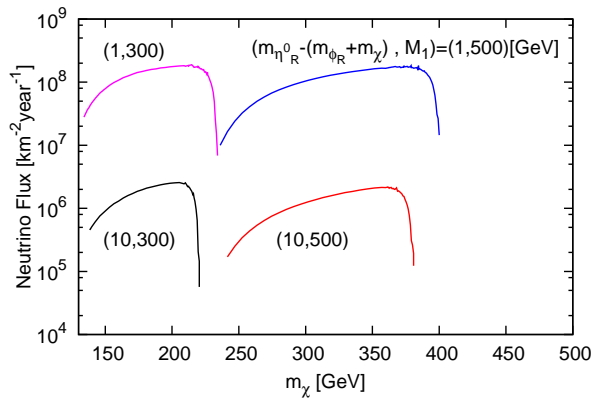}
	\caption{\label{nufluxreson} 
	\footnotesize 
	The neutrino flux from the Sun on the Earth
	against the $\chi$ DM mass.
	The flux is calculated from the reaction rate 
	$\Gamma(\nu) = C_A(\phi \chi \rightarrow N \nu)n_\phi n_\chi $, where
	the numbers $n_\phi $ and $n_\chi$ are obtained by solving the evolution equation  (\ref{Boltzsun}) numerically. 
We have used  four different  values of $m_{\eta^0_R} -(m_{\phi_R}+m_\chi)$ and $M_1$ :
$\left(m_{\eta^0_R} -(m_{\phi_R}+m_\chi), ~M_1\right)$=
(10 GeV, 300 GeV) (black curve), (1 GeV, 300 GeV) (magenta curve), (10 GeV, 500 GeV) (red curve), and (1 GeV, 500 GeV) (blue curve), respectively.}
\end{figure}

\section{conclusion}
In this paper our interest  has been directed at
an  indirect observation of multicomponent DM systems through
semi-annihilation processes of DMs,
because these processes are characteristic
of multicomponent DM systems.
In one-component DM systems of
a real scalar boson or of a Majorana fermion 
the monochromatic neutrino production by 
 DM annihilation is due to the chirality of the left-handed neutrino  strongly suppressed.
The suppression due to the chirality is absent 
when  DM is a complex scalar  boson or a  Dirac fermion.
In a multicomponent DM system, too, the neutrino production is unsuppressed
if it is an allowed process.

In this paper,
instead of performing a model independent investigation on 
 multicomponent DM systems we have first motivated
 the  existence of a multicomponent DM system
 by extending  the one-loop radiative seesaw model of Ma \cite{Ma:2006km}
to  remove  its shortcomings.
In the model of Ma \cite{Ma:2006km},
the lepton-number violating mass term of the 
 inert scalar doublet
has to be very small to obtain small neutrino masses.
This mass term originates from a lepton-number violating quartic 
scalar coupling,
the "$\lambda_5$ coupling", which is
$\mathcal{O}(10^{-5})$ to obtain small neutrino masses for $Y^\nu \sim 0.01$.
We therefore have considered an extension of  the model such that
this  lepton-number violating mass, too, is  radiatively generated.
Consequently, the seesaw mechanism occurs at the two-loop level
in the extended model \cite{Aoki:2013gzs}.
For this mechanism to work, we have  introduced 
a larger unbroken discrete symmetry, $Z_2\times Z_2$,
which implies that the model yields a
multicomponent DM system.
We emphasize that the 
multicomponent DM system is a consequence of the
 unbroken $Z_2\times Z_2$, which forbids the Dirac neutrino mass.

The DM annihilation processes can be classified to  three types;
standard annihilation, DM conversion and semi-annihilation.
We have assumed that the right-handed neutrino $N$
and two real bosons, $\chi$ and $\phi$, are DM particles,
and solved numerically the set of coupled Boltzmann equations.
It has turned out that the semi-annihilation effect 
for the heaviest dark matter is considerably
enhanced by the Boltzmann factor. 

We have computed the spin-independent cross section 
of the dark matter particles $\phi$ and $\chi$ off the nucleon.
(At the tree level there is no interaction of $N$ with the quarks.)
The quantity, which should be compared with the experimental
limits, is $\sigma_{\mathrm{ eff}}=(\sigma_{\chi} \Omega_{\chi}+\sigma_{\phi}\Omega_{\phi})/\Omega_{\rm total}$.
The predicted values of $\sigma_{\mathrm{eff}}$ have turned out to be
 slightly below the present limit given by LUX \cite{Akerib:2013tjd} for $m_{\chi} \gsim 150$ GeV.
Since the sensitivity of XENON1T \cite{Aprile:2012zx} will be 2 orders of magnitude
higher than that of  XENON100, 
the predicted area will be covered by XENON1T.
It should, however, be emphasized that the XENON1T experiment
alone cannot decide how many dark matter particles are present.
A clever choice of kinematical cuts at collider experiments could be used to explore a multi-component 
nature of DM \cite{Dienes:2014bka}.

As mentioned above, the monochromatic neutrino production 
by the semi-annihilation processes $\chi~N\to \nu_L~\phi$, etc., 
is not suppressed.
The time evolution of the number of the dark matter particles 
$n_{i}~(i=N, \phi,\chi$) in the Sun 
has been studied numerically to estimate
their values at the present time,
where we have set the capture rate for $N$ equal to zero.
Then we have calculated the reaction rate
$\Gamma(\nu)$ in the Sun, from which we have estimated 
the monochromatic neutrino flux coming from the Sun on the Earth
and hence
the monochromatic neutrino flux at the IceCube detector.
It turns out that the flux is very small compared with the current
IceCube sensitivity.
However,  the s-channel process of the semi-annihilation
 can be enhanced by a resonant effect:
 The enhanced signal is still 3 orders of magnetite smaller
 than the current IceCube sensitivity.
Nevertheless, the higher  the resolution of  energy and angle is,
the larger is the  chance for the observation of 
the monochromatic neutrino
 and hence of a multicomponent nature of DM.

\vspace*{5mm}
The work of M.~A.\ is supported in part by the Grant-in-Aid for Scientific 
Research (Grant No. 25400250 and No. 26105509), 
J.~K.\ is partially supported by the Grant-in-Aid for Scientific
Research (C) from the Japan Society for Promotion of Science (Grant No. 22540271), 
and H.~T\ is supported by Japan Society for the Promotion of Science (JSPS)　(Grant No. 13J05336).

\end{document}